\documentclass[12pt]{article}
\usepackage{slashed}
\usepackage{a4wide}
\usepackage{amsfonts}
\newcommand{\ft}[2]{{\textstyle\frac{#1}{#2}}}

\begin{document}
\setcounter{page}{0}
\thispagestyle{empty}
\begin{flushright}
KUL-TF/02-12\\
hep-th/0211192\\[4cm]
\end{flushright}
\begin{center}
{\Large \textsc{Solutions to the Massive HLW IIA Supergravity}}\\[2cm]
{\bf Jos Gheerardyn}\footnote{Aspirant FWO}\\
\vspace{1cm}
\textit{Instituut voor Theoretische Fysica, Katholieke Universiteit Leuven \break
Celestijnenlaan 200D, B-3001 Leuven, Belgium}\\[.5cm]
\texttt{Email : jos.gheerardyn@fys.kuleuven.ac.be}\\[4cm]
{\bf Abstract}\\[.5cm]
\begin{quote}
{\small We find new supersymmetric solutions of the massive supergravity theory which can be constructed by generalized
Scherk-Schwarz dimensional reduction of eleven dimensional supergravity, using the scaling symmetry of the
equations of motion. Firstly, we construct field configurations which solve the ten dimensional equations of motion by reducing on the radial direction of Ricci-flat cones. Secondly, we will extend this result to the supersymmetric case by performing a dimensional reduction along the flow of a homothetic Killing vector which is the Euler vector of the cone plus a boost.}
\end{quote}
\end{center}
\newpage
\section{Introduction}
In regular supergravities, the fields in the gauge multiplet are always massless. Massive theories are continuous deformations of the regular ones in which the gauge transformations and supersymmetries get extra dependence via a mass parameter $m$. Consequently, the equations of motion get extra terms linear and quadratic in this parameter. Some of the gauge fields (the St\"uckelberg fields) then can be gauged away, giving mass to other fields.

The prime example of these theories is Romans' IIA supergravity \cite{Romans:1986tz}. Its role in string theory was clarified after the discovery of D-branes \cite{Polchinski:1995mt}. It then became clear that this supergravity is actually the low energy limit of string theory in the background of a D8-brane. Such a brane solution was found in Romans' theory \cite{Polchinski:1996df,Bergshoeff:1996ui}, and the mass parameter was seen to be proportional to the charge of the D8.
    
It is a general property of massive supergravities that they admit domainwall solutions. Therefore, they are relevant to e.g. the domainwall/CFT correspondence \cite{Boonstra:1998mp,Behrndt:1999mk} and the Randall-Sundrum scenario \cite{Randall:1999ee,Randall:1999vf}.

A less known massive IIA theory is the one constructed by Howe, Lambert and West \cite{Howe:1998qt}. Although they constructed the theory by introducing a conformal spin connection, it is believed  \cite{Lavrinenko:1998qa} that the theory can also be constructed by performing a generalized Scherk-Schwarz reduction \cite{Scherk:1979zr} of the eleven dimensional equations of motion, using their scaling symmetry. The resulting massive theory does not have an action and has got no fundamental strings as the two-form potential (together with the dilaton) are St\"uckelberg fields. It was noted that this theory admits a de Sitter solution \cite{Lavrinenko:1998qa}. As only few other solutions are known \cite{Howe:1998qt,Chamblin:1999ea,Bergshoeff:2002nv,Chamblin}, it is the purpose of this paper to present a method of constructing (supersymmetric) solutions to that theory, using cones of special holonomy.

The plan of the paper is as follows. In section \ref{ansatz} we briefly remind how the massive theory can be constructed using generalised Scherk-Schwarz reduction. In section \ref{homothetic}, we point out that it is possible to perform this dimensional reduction in backgrounds admitting a homothetic Killing vector. Section \ref{ricciflat} comments on a specal kind of such backgrounds, namely Ricci-flat cones. Section \ref{susy} shows how we can extend this method to find backgrounds which preserve some fraction of supersymmetry and the final section \ref{conclusions} contains our conclusions. 
\section{Reduction Ansatz} \label{ansatz}
In this section, we will find the equations of motion and the supersymmetry rules by performing generalized
Scherk-Schwarz reduction of the eleven dimensional supergravity theory.

In regular Kaluza-Klein truncation on a circle, spacetime admits a $u(1)$ action generated by a Killing vector. Because all fields have to be periodic\footnote{Possibly modulo $\mathbb{Z}_2$ for fermionic fields.} on this circle, one can do a Fourier expansion of all fields. Dimensional reduction then boils down to retain only the zero modes. In that case, the remaining theory effectively does not depend on the extra dimension any more.

If the theory admits a global symmetry, it is sometimes possible to give the fields a very specific dependence on the circular coordinate, in such a way that the lower dimensional theory neither depends on the extra dimension. To be more concrete, when one is moving in the extra dimension, the fields undergo a symmetry transformation. This is called generalized Scherk-Schwarz reduction.
  
The equations of motion of eleven dimensional supergravity \cite{Cremmer:1978km} admit a rigid scaling symmetry
\begin{equation}\label{scaling}
e_\mu{}^a \rightarrow \lambda e_\mu{}^a \; , \; A_{\mu \nu \kappa} \rightarrow \lambda^3 A_{\mu \nu \kappa}\; , \;
\psi_\mu \rightarrow \lambda^{\frac{1}{2}} \psi_\mu \; ,
\end{equation}
and it can be used to construct a massive IIA supergravity \cite{Lavrinenko:1998qa}.
\subsection{Bosonic Fields}
Our conventions are explained in Appendix \ref{conventions}. The Ansatz for the Vielbein is taken to be to be
\begin{equation} \label{ansatze}
\hat{e}_{n}{}^{A}=e^{mz}\left(
 \begin{array}{cc}
 e^{\frac{1}{12}\phi}e_\mu{}^a & e^{-\frac{2}{3} \phi}A_\mu \\
 0& e^{-\frac{2}{3} \phi}
 \end{array} \right)
 \quad
\hat{e}_{A}{}^{n}=e^{-mz}\left(
 \begin{array}{cc}
 e^{-\frac{1}{12}\phi}e_a {}^\mu & - e^{-\frac{1}{12} \phi}A_a \\
 0& e^{\frac{2}{3} \phi}
 \end{array} \right)\; , 
\end{equation}
while the three form must depend on $z$ in the following way
\begin{equation}
\hat{A}_3=e^{3mz}(A_3+A_2 \wedge dz)\; .
\end{equation}
With this three form, we define the ten dimensional field strengths to be
\begin{eqnarray}
F_2&=& dA_1\; ,\nonumber\\
F_3&=& dA_2-3mA_3\; ,\nonumber\\
F_4&=&dA_3-dA_2\wedge A_1+3mA_3\wedge A_1\; .
\end{eqnarray}
\subsection{Fermionic Fields}
The Ansatz for the gravitino is taken in such a way that the supersymmetry of the lower dimensional Vielbein has the usual form
\begin{equation}\label{viel}
\delta e_\mu {}^a= \bar{\epsilon} \gamma^a \psi_\mu\; .
\end{equation}
Taking into account the scaling property (\ref{scaling}) together with the fact that we are reducing on flat indices, we take
\begin{equation}
\hat{\psi}_a=e^{-\frac{1}{24} \phi -\frac{1}{2}mz}(\psi_a-\ft18 \gamma_a \lambda)\; , \;
\hat{\psi}_{i}=e^{-\frac{1}{24} \phi -\frac{1}{2}mz}\gamma_{11} \lambda\; ,
\end{equation}
while the parameter for supersymmetry has to satisfy
\begin{equation}\label{ansatzsusy}
\hat{\epsilon}=e^{\frac{1}{24} \phi +\frac{1}{2}mz}\epsilon\; .
\end{equation}
Using this Ansatze, we can derive the equations of motion and the supersymmetry transformation rules, which are given in the Appendix \ref{theory}. Note that all ten dimensional fields now only depend on the ten coordinates $x^\mu$.
\section{Homothetic Killing Vectors}\label{homothetic}
In this paper, we will consider purely gravitational (bosonic) solutions of M theory. As a consequence of the equations of motion, these solutions are Ricci-flat.
If the metric admits a homothetic Killing vector $k$, there will exist a coordinate system in which it will satisfy the reduction Ansatz. This can easily be seen by choosing coordinates adapted to $k$ ($k=\partial_z$), as the Lie derivative then reduces to
\begin{equation}
\mathcal{L}_k \hat{g}_{mn}=k^p \partial_p \hat{g}_{mn}=2m\hat{g}_{mn}\; ,
\end{equation}
which is solved by
\begin{equation}\label{solg}
\hat{g}=e^{2m z} \hat{h}(x)\; .
\end{equation}
As the Ansatz for the Vielbein (\ref{ansatze}) implies that the metric reads
\begin{equation} \label{ansatzg}
\hat{g}=e^{2mz}\left( e^{\frac16 \phi} g_{\mu \nu}dx^\mu dx^\nu+e^{-\frac43 \phi}(dz+A_\mu dx^\mu)^2 \right)\; ,
\end{equation}
the eleven dimensional metric (\ref{solg}) written in the coordinates adapted to $k$ will yield a solution of the massive ten dimensional theory.

If this classical M theory field configuration moreover preserves $\mathcal{N}$ supersymmetries, it will admit $\mathcal{N}$ parallel spinors $\hat{\epsilon}$. Looking at the Ansatz (\ref{ansatzsusy}) for the parameters of supersymmetry, we see that parallel spinors satisfying the Ansatz are solutions of
\begin{equation}
\partial_z \hat{\epsilon}=\ft m2 \hat{\epsilon}\; .
\end{equation}
Written in a coordinate invariant way, we find that the dimenionally reduced solution will preserve as many supersymmetries as there are eleven dimensional spinors satisfying
\begin{equation}
\mathcal{L}_k \hat{\epsilon}=k^m \nabla_m \hat{\epsilon}+\ft14 \partial_m k_n \Gamma^{mn} \hat{\epsilon}=\ft m2 \hat{\epsilon} \; .
\end{equation}
This condition can even be simplified to
\begin{equation}\label{pressusy}
\partial_m k_n \Gamma^{mn} \hat{\epsilon}=2m \hat{\epsilon}\; ,
\end{equation} 
as the spinors are parallel.

If we would also consider solutions with a three form differing from zero, it would satisfy the Ansatz if
\begin{equation}
\mathcal{L}_{k} \hat{A}_3=3m\hat{A}_3 \; .
\end{equation}

In conclusion, a gravitational solution of classical M theory yields a solution of the massive theory if it admits a homothetic Killing vector. The number of preserved supersymmetries of the ten dimensional field configuration equals the number of eleven dimensional parallel spinors satisfying (\ref{pressusy}).
\section{Ricci-flat Cones}\label{ricciflat}
To clarify the previous discussion, we will now perform a generelaized Scherk-Schwarz reduction of eleven dimensional cones on their Euler vector, yielding non-supersymmetric solutions if the massive IIA theory.

If a manifold admits a homothetic Killing vector $k$ which is hypersurface orthogonal, i.e. there exist a function $f$ such that
\begin{equation}
k_\mu=\partial_\mu f \; ,
\end{equation}
then there is a set of coordinates in which the metric can be written as a cone \cite{Gibbons:1998xa}. The vectorfield $k$ is then called the Euler vector. 

We now start our construction by taking a solution of eleven dimensional supergravity which is $11-d$ dimensional Minkowski space times a $d$ dimensional Ricci-flat Riemannian cone and write Minkowski space as a cone over de Sitter space.
\begin{equation}\label{g2}
\hat{g}=dR^2+R^2ds^2_{dS}+dr^2+r^2 ds^2_{d-1}\; .
\end{equation}
Now, $k=R\partial_R+r\partial_r$ is a homothetic Killing vector which is hypersurface orthogonal.
If we change coordinates to
\begin{equation}
r=e^z \cos \rho\, , \, \, R=e^z \sin \rho\; , \;\; \rho \in [-\ft{\pi}2,+\ft{\pi}2] \; ,
\end{equation}
the homothetic Killing vector becomes $k=\partial_z$ and the metric reads
\begin{equation}
\hat{g}=e^{2z}\left(dz^2+d\rho^2+\sin^2 \rho ds_{dS}^2+\cos^2 \rho ds^2_{d-1} \right) \; .
\end{equation}
From the reduction Ansatz (\ref{ansatzg}), we can directly read of that 
\begin{equation}
g=d\rho^2+\sin^2 \rho ds_{dS}^2+\cos^2 \rho ds^2_{d-1}\; , \; \; m=1 \; ,
\end{equation}
and all other fields zero, is a solution of the massive supergravity. We can find solutions for any value of the mass parameter by reducing on $mk$.

A large set of Ricci-flat Riemannian cones are cones of special holonomy \cite{Acharya:1998db}. In this reference, a lot of examples can also be found. 
\section{Supersymmetric Reductions}\label{susy}
In this section, we will list solutions of the massive theory which are reductions of special holonomy cones (supplied with some extra flat directions) and which preserve some fraction of supersymmetry. The vector field we will use to perform the dimensional reduction will now be the Euler vector from the previous section plus a boost in the flat directions. Therefore, the vector field will not be hypersurface orthogonal anymore and we will be able to preserve supersymmetry during the reduction process. 
\subsection{Reduction of Flat Space}
We will start with the easiest case of the supersymmetric reduction of Minkowski space, where the ten dimensional solution will preserve half of the supersymmetries.
Flat space admits a homothetic hypersurface orthogonal Killing vector $K$ which can be written in natural coordinates as 
\begin{equation}
K=x^m \partial_m \; , \; \; \mathcal{L}_K\eta_{mn}=2\eta_{mn}\; ,
\end{equation}
implying that Minkowski space is a cone over de Sitter space and $K$ is the Euler vector.
 Noting that all $so(1,10)$ rotations induce Killing vectors $l$ of Minkowski space, the vector field 
 $k=K+l$ still is homothetic, but not hypersurface orthogonal anymore. The important observation to be able 
 to construct supersymmetric solutions is that we can choose $l$ such that 
\begin{equation}
\mathcal{L}_k\hat{\epsilon}=\mathcal{L}_l \hat{\epsilon}=\ft12 \hat{\epsilon}\; ,
\end{equation}
for some parallel spinors $\hat{\epsilon}$ of eleven dimensional Minkowski space. The spinors satisfying this condition reduce to supersymmetries of the ten dimensional solution.

We can write the Killing vector $l$ as 
\begin{equation}
l=B_m{}^n x^m \partial_n \; .
\end{equation}
In that case, the condition for supersymmetry (\ref{pressusy}) reads
\begin{equation}
\slashed{B}\hat{\epsilon}=2 \hat{\epsilon}\; , \; \; \slashed{B}=B_m{}^n \eta_{np}\Gamma^{mp}\; .
\end{equation}
As the spinors have to have real eigenvalues of $\slashed{B}$, $l$ has to be a boost, and it is always possible to take it in the $(x^0,x^1)$ plane. Because $\Gamma^{01}$ squares to one and is traceless, half of its eigenvalues are $1$, and the other half are $-1$. Therefore, we take $l$ to be 
\begin{equation}
l=x^1\partial_0+x^0\partial_1 \; .
\end{equation}
As a consequence, the homothetic Killing vector we want to reduce on, reads
\begin{equation}
k=(x^0+x^1)\partial_0+(x^0+x^1)\partial_1+x^a
\partial_a=(x^0+x^1)\partial_0+(x^0+x^1)\partial_1+r\partial_r\; ,
\end{equation}
where $r=\sqrt{x^ax^a}$ and $a=2,\dots,10$. Firstly, we write the flat metric as
\begin{equation}
\hat{g}=-(dx^0)^2+(dx^1)^2+\Big(dr^2+r^2d\Omega_8^2\Big)\; ,
\end{equation}
where $d\Omega_8^2$ is the natural metric on the 8-sphere. We can now choose new coordinates\footnote{The Jacobian for this coordinate transformation becomes singular for $y_2=0$.} which are adapted to the vector field $k=\partial_z$.
\begin{eqnarray}
x^0&=&\ft12 y_2\left( e^{2z}+e^{-2y_1} \right)\; ,\nonumber \\
x^1&=&\ft12y_2 \left( e^{2z}-e^{-2y_1} \right)\; ,\nonumber\\
r&=&e^{z-y_1}\; .
\end{eqnarray}
In these new coordinates, the eleven dimensional metric reads
\begin{eqnarray} \label{redmink}
\hat{g}&=&e^{2(z-y_1)} \Big( (dz-(1-2y_2^2)dy_1-y_2dy_2)^2\nonumber\\&&+4y_2^2(1-y_2^2)dy_1^2-(1+y_2^2)dy_2^2+4y_2^3dy_1dy_2+d\Omega_8^2 \Big)\; .
\end{eqnarray}
From the reduction Ansatz (\ref{ansatzg}), we can read of the ten dimensional field configuration. 
The ten dimensional solution is ($m=1$)
\begin{eqnarray}
g&=&e^{-\frac94y_1}\left( 4y_2^2(1-y_2^2)dy_1^2-(1+y_2^2)dy_2^2+4y_2^3dy_1dy_2+d\Omega_8^2 \right) ,\nonumber\\
A_{y_1}&=&-(1-2y_2^2) \; \;  , \; \; \;
A_{y_2}=-y_2\; ,\nonumber\\
\phi&=&\ft32 y_1\; .
\end{eqnarray}
This solution preserves $1/2$ of the supersymmetry. 
\subsection{Reductions on Special Holonomy Cones}
To find solutions with less supersymmetry, we will start from an eleven dimensional configuration which is the product of a Riemannian simply connected special holonomy manifold and flat Minkowski space. Firstly, we briefly comment on these eleven dimensional solutions, and secondly we find the appropriate homothetic Killing vector to perform a supersymmetric reduction. 
\subsubsection{Parallel Spinors}
If we use Ricci-flat manifolds as (purely gravitational) solutions of classical M theory, the condition for them to preserve some supersymmetry is that the supersymmetry variation of the gravitino is zero.
\begin{equation} \label{varpsi}
\delta \hat{\psi}_m=\nabla_m \hat{\epsilon}=0\; .
\end{equation}
If the solution is the product of flat space and a special holonomy manifold, the parallel spinors are
products of solutions of (\ref{varpsi}) on both spaces. Counting these
gives the number of preserved supersymmetries.
\begin{center}
\begin{tabular}[t]{|c|c|c|c|}
\hline
Dimension  & Manifold & Holonomy & Parallel Spinors\\
\hline 
\hline
$4n$&Calabi-Yau&$su(2n)$&$(2,0)$\\
$4n+2$&Calabi-Yau&$su(2n+1)$&$(1,1)$\\
$4n$&Hyperk\"ahler&$sp(2n)$&$(n+1,0)$\\
7&Exceptional&$G_2$&$1$\\
8&Exceptional&$spin(7)$&$(1,0)$\\
\hline
\end{tabular} 
\end{center} 
\subsubsection{Solutions}
The eleven dimensional solution we start with reads
\begin{equation}
\hat{g}=-(dx^0)^2+(dx^1)^2+dr_1^2+r_1^2d\Omega_{n-2}^2+dr_2^2+r_2^2 ds_{9-n}^2\; ,
\end{equation}
where the last two terms are the metric on the $10-n$ dimensional special holonomy cone. We first perform a coordinate transformation
\begin{equation}
r_1=r \cos \alpha \; \; , \; \; \; r_2=r\sin \alpha \; ; \; \; \alpha \in [0,\ft{\pi}2] \; , \; n>2 \; | \;\alpha \in [-\ft{\pi}2,\ft{\pi}2] \; , \; n=2 \; .
\end{equation}
The metric in these new coordinates now reads
\begin{equation}
\hat{g}=-(dx^0)^2+(dx^1)^2+dr^2+r^2 \Big(d\alpha^2 + \cos^2 \alpha d\Omega_{n-2}^2+\sin^2 \alpha ds_{9-n}^2\Big)\; ,
\end{equation}
while the homothetic Killing vector we will use for the reduction reads
\begin{equation}
k=(x^0+x^1)\partial_0+(x^0+x^1)\partial_1+r\partial_r\; .
\end{equation}
Comparing with (\ref{redmink}), we see that we only have to substitute the metric on the 8-sphere by
\begin{equation}
d\Omega_8^2 \to  d\alpha^2 + \cos^2 \alpha d\Omega_{n-2}^2+\sin^2 \alpha ds_{9-n}^2\; .
\end{equation}
The number of preserved supersymmetries is now half of the number preserved by the eleven dimensional solution.
We list this number $\mathcal{N}$ in the following table together with the dimension of the special holonomy cone. Of course, it is always possible to
consider products of different manifolds.
\begin{center}
\begin{tabular}[t]{|c|c|c|}
\hline
Dimension&Manifold& $\mathcal{N}$\\
\hline \hline
8&Spin(7)&1\\
8&$CY_4$&2\\
8&$HK_2$&3\\
7&$G_2$&2\\
6&$CY_3$&4\\
4&$CY_2= HK_1$&8\\
\hline
\end{tabular}
\end{center}
\section{Conclusions}\label{conclusions}
In this paper, we have constructed several sets of solutions of the massive supergravity first built by
Howe, Lambert and West. In the first part of the paper, we have found non-supersymmetric solutions by reducing on the
radial direction of cones. In the second part of this paper, we have built supersymmetric solutions by
using cones of special holonomy and by reducing on a homothetic Killing vector which is the sum of a scale
transformation and a boost.
\section*{Acknowledgments}
We would like to thank B. Janssen, P. Meessen, F. Roose and A. Van Proeyen
for useful discussions. This work has been partially supported by the FWO-Vlaanderen and the RTN Network
\emph{The quantum structure of spacetime and the geometric nature of fundamental
interactions} HPRN-CT-2000-00131.
\appendix
\section{Conventions}\label{conventions}
The field content of eleven dimensional supergravity is a Vielbein $\hat{e}_m^A$ (metric $\hat{g}$), a three form $\hat{A}_3$ and a gravitino $\hat{\psi}_m$. Eleven dimensional curved indices are denoted by $m,n,\dots$, while flat labels are $A,B,\dots$. The parameter for supersymmetries are $\mathbf{32}$ Majorana spinors $\hat{\epsilon}$. The direction along which the dimensional reduction is performed is denoted by $z$ for a curved and $i$ for a flat direction. The metric is taken to be mostly plus and a covariant derivative on a spinor is 
\begin{equation}
\nabla_m \hat{\epsilon}=\partial_m \hat{\epsilon}+\ft14 \slashed{\omega}_m \hat{\epsilon}=\partial_m \hat{\epsilon}+\ft14 \omega_m{}^{ab} \gamma_{ab} \hat{\epsilon}\; .
\end{equation}
Eleven dimensional gamma matrices are $\Gamma^m$ and they satisfy
\begin{equation}
\{ \Gamma_m , \Gamma_n \}=2g_{mn}.
\end{equation}
Ten dimensional curved indices are $\mu, \nu, \dots$ while flat labels are $a,b,\dots$. Ten dimensional gamma matrices are $\gamma^\mu$. Covariant derivatives are written as $D_\mu$. 

\noindent
Brackets $(m n)$ denote symmetrization in $m,n$ with unit weight while $[m,n]$ means anti-symmetrisation.
\section{Equations of Motion and Supersymmetry} \label{theory}
\subsection{Equations of Motion}
If we substitute the Ansatz in the eleven dimensional equations of motion, we find that all $z$-dependence disappears. The equations of motion for the bosonic fields then become \cite{Lavrinenko:1998qa,Howe:1998qt}
\begin{displaymath}
\Box\phi = -\ft38 e^{-\ft32\phi}\, F_2^2 +\ft1{12}
e^\phi\, F_3^2 -\ft1{96} e^{-\ft12\phi}\, F_4^2 +\ft{27}2 m^2\,
A_\mu\, A^\mu + 9m\, A^\mu\partial_\mu\phi -\ft32 m\,
D_\mu{\cal A}^\mu
\end{displaymath}
\begin{eqnarray}
R_{\mu\nu} -\ft12 R\, g_{\mu\nu}&=& \ft12(\partial_\mu\phi\,
\partial_\nu\phi -\ft12 (\partial\phi)^2\, g_{\mu\nu}) 
+\ft12 e^{-\ft32\phi}\, (F_{\mu\rho}\, F_\nu{}^\rho 
-\ft14F_2^2\, g_{\mu\nu})\nonumber\\
&&+\ft1{4} e^\phi\, (F_{\mu\rho\sigma}\,
F_\nu{}^{\rho\sigma} -\ft16 F_3^2\, g_{\mu\nu}) + 
\ft1{12} e^{-\ft12\phi}\, (F_{\mu\rho\sigma\lambda}\,
F_\nu{}^{\rho\sigma\lambda} -\ft18 F_4^2\, g_{\mu\nu})\nonumber\\
&&
-9m^2\, (A_\mu\, A_\nu + 4A_\rho\, A^\rho\, g_{\mu\nu}) - 36m^2\, e^{\ft32\phi}\, g_{\mu\nu} \nonumber\\
&&-\ft92 m\, (D_\mu A_\nu +D_\nu A_\mu
-2D_\rho A^\rho\, g_{\mu\nu})\nonumber\\
&&+\ft34m\, (A_\mu\partial_\nu\phi + A_\nu\partial_\mu\phi -A^\rho\partial_\rho \phi\, g_{\mu\nu})\nonumber\\
D_\nu(e^{-\ft32\phi}\, F_\mu{}^\nu) &=& 12m\partial_\mu\phi
+18m^2\, A_\mu + 9m\, e^{-\ft32\phi}\, A^\nu\, F_{\mu\nu} -\ft16 e^{-\ft12\phi}\, F_{\mu\nu\rho\sigma}\nonumber
F^{\nu\rho\sigma}\nonumber\\ 
D^{\sigma}(e^{-\ft12\phi}F_{\mu\nu\rho\sigma})&=& -6m\,e^{\phi}\,
F_{\mu\nu\rho}+6m\, e^{-\ft12\phi}\, A^{\sigma}F_{\mu\nu\rho\sigma}
-\ft1{144}\epsilon_{\mu\nu\rho\sigma_1\dots\sigma_7}\,
F^{\sigma_1\dots\sigma_4}\,F^{\sigma_5\sigma_6\sigma_7}\nonumber\\
D^{\sigma}(e^{\phi}F_{\mu\nu\sigma})&=& 6m\,e^{\phi}\, A^\sigma \, 
F_{\mu\nu\sigma}+\ft12\, e^{-\ft12\phi}\, F_{\mu\nu\sigma\rho}\, 
F^{\sigma\rho} + \ft1{1152}
\epsilon_{\mu\nu\rho_1\dots\rho_8}\, F^{\rho_1\dots\rho_4}
F^{\rho_5\dots\rho_8}\nonumber\\
\end{eqnarray}
Note that these equations of motion cannot be derived from an action.
\subsection{Supersymmetry}
If we use the Ansatz to deduce the supersymmetry transformation rules, we find \cite{Bergshoeff:2002nv}
\begin{eqnarray}
\delta \lambda&=& \ft12 m e^{\frac{3}{4} \phi} \gamma_{11} \epsilon-\ft18 e^{-\frac{3}{4} \phi} \slashed{F}_2
\gamma_{11} \epsilon-\ft13 \slashed{\mathfrak{D}} \phi
\epsilon-\ft{1}{144} e^{-\frac{1}{4} \phi} \slashed{F}_4 \epsilon+\ft{1}{18} e^{\frac{1}{2} \phi}
\slashed{F}_3 \gamma_{11} \epsilon \nonumber\\
\delta \psi_a&=&D_a \epsilon+\ft{9}{16}m e^{\frac{3}{4} \phi} \gamma_a \gamma_{11}
\epsilon-m\ft{9}{16} \gamma_a \slashed{A}_1\epsilon-\ft{1}{64} e^{-\frac{3}{4} \phi}({}_a \slashed{F}_2-14
\slashed{F}_{(2)a})\gamma_{11} \epsilon \nonumber\\&&+\ft{1}{48} e^{\frac{1}{2} \phi}(9\slashed{F}_{(3)a}-{}_a
\slashed{F}_3) \gamma_{11} \epsilon+\ft{1}{128} e^{-\frac{1}{4} \phi}(\ft{20}{3} \slashed{F}_{(4)a}-{}_a
\slashed{F}_4)\epsilon
\end{eqnarray}
with
\begin{equation}
\mathfrak{D}_a \phi=\partial_a \phi+\ft32 m A_a
\end{equation}

\providecommand{\href}[2]{#2}\begingroup\raggedright
\end{document}